\documentclass[aps,pra,twocolumn,showpacs,superscriptaddress,tightenlines,preprintnumbers,sort&compass,amsmath,amssymb,floatfix,nofootinbib]{revtex4}

\usepackage{hyperref}
\usepackage{amsmath}
\usepackage{graphicx}
\usepackage{amssymb}
\usepackage{epsfig}
\usepackage{amsfonts}
\usepackage{color}

\usepackage[utf8x]{inputenc}
\usepackage[T1]{fontenc}

 \def\>{\rangle}

\newcommand{\p}{\mathbf{p}}
\newcommand{\s}{\mathbf{s}}

\newcommand{\ket}[1]{\mbox{$|#1\rangle$}}

\begin{document}

\title{Two methods for measuring Bell nonlocality via local unitary invariants of two-qubit systems in Hong-Ou-Mandel interferometers}

\author{Karol Bartkiewicz}
\email{bark@amu.edu.pl} \affiliation{Faculty of Physics, Adam
Mickiewicz University, PL-61-614 Pozna\'n, Poland}
\affiliation{RCPTM, Joint Laboratory of Optics of Palack\'y
University and Institute of Physics of Academy of Sciences of the
Czech Republic, 17. listopadu 12, 772 07 Olomouc, Czech Republic }

\author{Grzegorz Chimczak}
\affiliation{Faculty of Physics, Adam
Mickiewicz University, PL-61-614 Pozna\'n, Poland}

\begin{abstract}
We describe a direct method to experimentally determine
 local two-qubit invariants  by 
performing interferometric measurements on multiple
copies of a given two-qubit state. We use this framework to analyze two different
kinds of two-qubit invariants of Makhlin and Jing et. al.
These invariants allow 
to fully reconstruct any two-qubit state up to local unitaries.
We demonstrate that measuring 3 invariants is sufficient to find, e.g.,  
the optimal Bell inequality violation. These invariants can be measured 
with local or nonlocal measurements.
We show that the nonlocal strategy that follows from
Makhlin's invariants is more resource-efficient than
local strategy following from the invariants of Jing et al.
To measure all of the Makhlin's invariants 
directly one needs to use both
two-qubit singlet and three-qubit W-state projections 
on multiple copies of the two-qubit state. This problem is 
equivalent to a cordinate system handness measurement.
We demonstrate that these 3-qubit measurements can be performed
by utilizing Hong-Ou-Mandel interference which gives significant
speedup in comparison to the classical handness measurement.
Finally, we point to potential application of our results 
in quantum secret sharing.
\end{abstract}

\pacs{03.67.Mn, 42.50.Dv}


\date{\today}
\maketitle

\section{Introduction}

Local unitary invariants are fundamental quantities that do not change
after performing local unitary transformations on subsystems of a composite quantum 
system~\cite{Grassl1998,Kempe1999,Makhlin2002,King2007,Jing2015}. 
In a way they are similar
to constants of motion in classical mechanics, which 
remain unchanged under some transformations performed
locally on  coordinate systems of its parts.
The invariants are proved to be a useful and powerful mathematical tool 
that can be applied in designing and analyzing quantum gates~\cite{Makhlin2002,Koponen2005}, quantum error correction~\cite{Ranis2000} and
for measuring quantum correlations~\cite{Osterloh12,Carteret05PRL,Bartkiewicz15a,Bartkiewicz15b}.
In this paper, we focus on a two-qubit case, which is especially important
for practical applications as two-qubit correlations are necessary for 
performing various quantum information processing and quantum 
communications tasks that rely on quantum entanglement~\cite{Schroedinger35,EPR35,Horodecki09RMP}.
These applications include, e.g., dense coding~\cite{Bennett92PRL}, quantum teleportation~\cite{Bennett93PRL},
entanglement swapping~\cite{Zukowski93PRL}, entanglement-based quantum key distribution~\cite{Ekert91PRL,Gisin02RMP}, 
quantum repeaters~\cite{Dur99PRA}, 
quantum nondemolition photon detection~\cite{Bula13QND,Scott13AMP} used
for qubit amplification~\cite{Gisin10PRL}.
Moreover, the quantum correlations can be interpreted as a manifestation of
nonlocality and detected by breaking the Bell-Clauser-Horne-Shimony-Holt 
inequality~\cite{Bell64,Bell,CHSH69PRL,Brunner14RMP}.

Here, we demonstrate that local invariants are not only 
a convenient tool to analyze these phenomena, but also they can be used
to design new experiments for measuring quantum correlations 
and other nonlinear properties of quantum states (like, e.g., nonlocality).
For this purpose we focus on two sets of local unitary invariants, i.e., Makhlin's invariants $I$ 
from Ref.~\cite{Makhlin2002} 
and invariants of Jing et al. $J$ form Ref.~\cite{Jing2015}. We show that all the investigated invariants can be expressed 
as expected values of measurements performed on multiple copies 
of a given two-qubit system. Hence, each invariant can be expressed as 
a combination of measurements with outcomes valued $\pm1.$
We group these composite measurements in three categories, i.e., local chained,
local looped, and nonlocal
measurements shown in Figs.~\ref{fig:chained}-\ref{fig:nonlocal}.
The local measurements are invariant under local unitaries
and their prime element is a singlet projection, which
is naturally implemented in linear optical systems 
by measuring anticoalescence rate of photons that interfered 
on a balanced beam splitter, i.e., by measuring Hong-Ou-Mandel (HOM) interference~\cite{HOM}.
Similar composite HOM measurements were used in several experimental and theoretical 
works related to detecting and measuring, e.g., quantum entanglement, quantum discord, 
purity of quantum states, and performing optimal quantum tomography or measuring spectra
of density matrices
(see, e.g., Refs. ~\cite{Bovino05PRL,Walborn06Nat,Jin2012,Bartkiewicz13corr,Bartkiewicz13PRA_CHSH,Rudnicki11PRL,Rudnicki12PRA,
Bartkiewicz13sfid,Tanaka13PRA,Tanaka14PRA,Bartkiewicz15a,Bartkiewicz15b,tomo1,tomo2,Bartkiewicz17fef,Lemr16collect,Bartkiewicz17hom}).
The vast subject of multiphoton interferometry is reviewed in Ref.~\cite{Pan12RMP}.
Here, we show that some of the most complex Makhlin's invariants~\cite{Makhlin2002}
can be expressed by projections on 3 particle $W$-states, which do not
exhibit bipartite entanglement~\cite{Dur00PRA}. The results of such measurements 
can be interpreted as measuring handness of a coordinate system formed 
by three Bloch vectors. 
We demonstrate that even by using  projections on maximally entangled two-qubit states  
it is possible to perform the handness measurement much faster than by using  the classical approach to the problem based on separable 
single-qubit projections.

In this paper, we describe two alternative ways of performing a test of
Bell-Clauser-Horne-Shimony-Holt  (Bell-CHSH) inequality violation ~\cite{CHSH69PRL} 
to the approaches known from the literature~\cite{CHSH69PRL,Aspect81PRL,Aspect82aPRL,Aspect82bPRL,
Horodecki95PLA,Horodecki96aPLA,Horst13PRA,Bartkiewicz13PRA_CHSH,Bartkiewicz17fef,Bartkiewicz17hom,Brunner14RMP,Hensen15Nature,Shalm15PRL}. 
Each of these methods is related to a different set of invariants and allows to directly test 
the optimal Bell-CHSH inequality and quantify the level of its violation. We also show 
that the presented interferometers can be also used for  measuring the fully entangled 
fraction~\cite{Bennett96}, which is useful for estimating the fidelity of many 
entanglement-based quantum information protocols (see, e.g., Refs.~\cite{Bennett96,Horodecki96PRA,Grondalski02PLA,Badziag00PRA,Bartkiewicz17fef}).

This paper is organized as follows: 
In Sec.~II, we establish the theoretical framework to be used for expressing 
the invariants $I$ and $J$ in terms of quantities which are measurable
via HOM interference. In Sec.~III, the Makhlin's and Jing's et al. invariants
are defined via experimentally-accessible state projections. 
In Sec.~IV we describe two new approaches towards measuring Bell-CHSH nonlocality
and other quantities based on  invariants $I$ and $J$, e.g., fully-entangled fraction. 
We conclude in Sec.~V.

\section{Theoretical framework}
\subsection{Two-qubit density matrix}
A two-qubit density matrix can be represented in standard Hilbert-Schmidt form
using Einstein's summation convention as
\begin{equation}
\hat\rho_{a,b}=( \tfrac{1}{4}\hat\sigma_0^{\otimes 2}+ s_i\hat\sigma_i\otimes\hat\sigma_0 + p_i\hat\sigma_0\otimes\hat\sigma_i  +\beta_{i,j}\hat\sigma_i\otimes\hat\sigma_j )_{a,b},
\end{equation} 
where in the case of photonic polarization qubits the Pauli matrices can
be expressed in terms of projections on  horizontal $\ket{H},$ vertical $\ket{V},$
diagonal $\ket{D},$ antidiagonal $\ket{A},$ left-circular $\ket{L}$, and
right-circular $\ket{R}$ single-photon  polarization states, i.e.,
as $\hat\sigma_0=|H\rangle\langle H|+|V\rangle\langle V|,$  
$\hat\sigma_1=|D\rangle\langle D|-|A\rangle\langle A|\equiv \hat\sigma_x,$
$\hat\sigma_2=|L\rangle\langle L|-|R\rangle\langle R|\equiv \hat\sigma_y,$ and
$\hat\sigma_3=|H\rangle\langle H|-|V\rangle\langle V|\equiv \hat\sigma_z.$
The photons observed individually have  Bloch vectors $\mathbf{s}$ and $\mathbf{p}$
for subsystems in modes $a$ and $b$, respectively. The correlations between the subsystems are 
described by matrix $\hat\beta$. 

\subsection{Singlet projections}
Projections on singlet state are often implemented in studying 
quantum aspects of polarization-encoded
two-qubit state and as an element of such quantum information processing task as, e.g., 
quantum teleportation, entanglement swapping, and dense coding etc.
The singlet projection can be implemented by a balanced
beam splitter (BS), which performs the following
operations on the Bell basis states [i.e., $|\Psi^\pm\rangle=\tfrac{1}{\sqrt{2}}(|HV\rangle\pm|VH\rangle)$
and $|\Phi^\pm\rangle=\tfrac{1}{\sqrt{2}}(|HH\rangle\pm|VV\rangle)$]
of pairs of photons in spatial modes $k$ and $l$
\begin{eqnarray}
 \hat U_{\rm BS} \ket{\Psi^{-}}_{k,l} &=& -\ket{\Psi^{-}}_{k,l}= -\tfrac{1}{\sqrt{2}}(\ket{H,V}-\ket{V,H})_{k,l}\nonumber \\
 \hat U_{\rm BS} \ket{\Psi^{+}}_{k,l} &=& \tfrac{1}{\sqrt{2}}(\ket{HV,0}-\ket{0,HV})_{k,l}\label{eq:BS}\\
 \hat U_{\rm BS} \ket{\Phi^{\pm}}_{k,l} &=&
  \tfrac{1}{2}(\ket{2H,0}-\ket{0,2H})_{k,l}\nonumber\\
&&  \pm \tfrac{1}{2}(\ket{2V,0}-\ket{0,2V})_{k,l},\nonumber
\end{eqnarray}
where   $\ket{0}$  is the vacuum, $\ket{V,H}_{k,l}=\hat b_{k,V}^{\dagger}\hat b_{l,H}^{\dagger}\ket{0,0}$, and $\ket{2V,0}_{k,l}=\tfrac1{\sqrt{2}}
\hat b_{k,V}^{\dagger2}\ket{0,0}_{k,l}$, etc.
These transformations can be derived in the Heisenberg picture with
input and output annihilation operators for polarization $p=H,V$ are
 $\hat a_{k,p},$ $\hat a_{l,p}$ and $\hat b_{k,p},$ $\hat b_{l,p},$ respectively.
For this BS the input-output relations read as $\hat a_{k,p}=(\hat b_{k,p}+\hat b_{l,p})/\sqrt{2}$ and $\hat a_{l,p}=(\hat b_{k,p}-\hat b_{l,p})/\sqrt{2}.$ Thus, if we detect at the same time one photon in each output 
of the BS, we perform the singlet projection.
Similarly, if we detect two photons of orthogonal polarizations 
in a single output mode of the BS, we perform the $\ket{\Psi^+}_{k,l}$ state projection.
In the latter case, one usually uses polarizing beam splitters (PBS).

\begin{figure}
\includegraphics[width=0.6\linewidth]{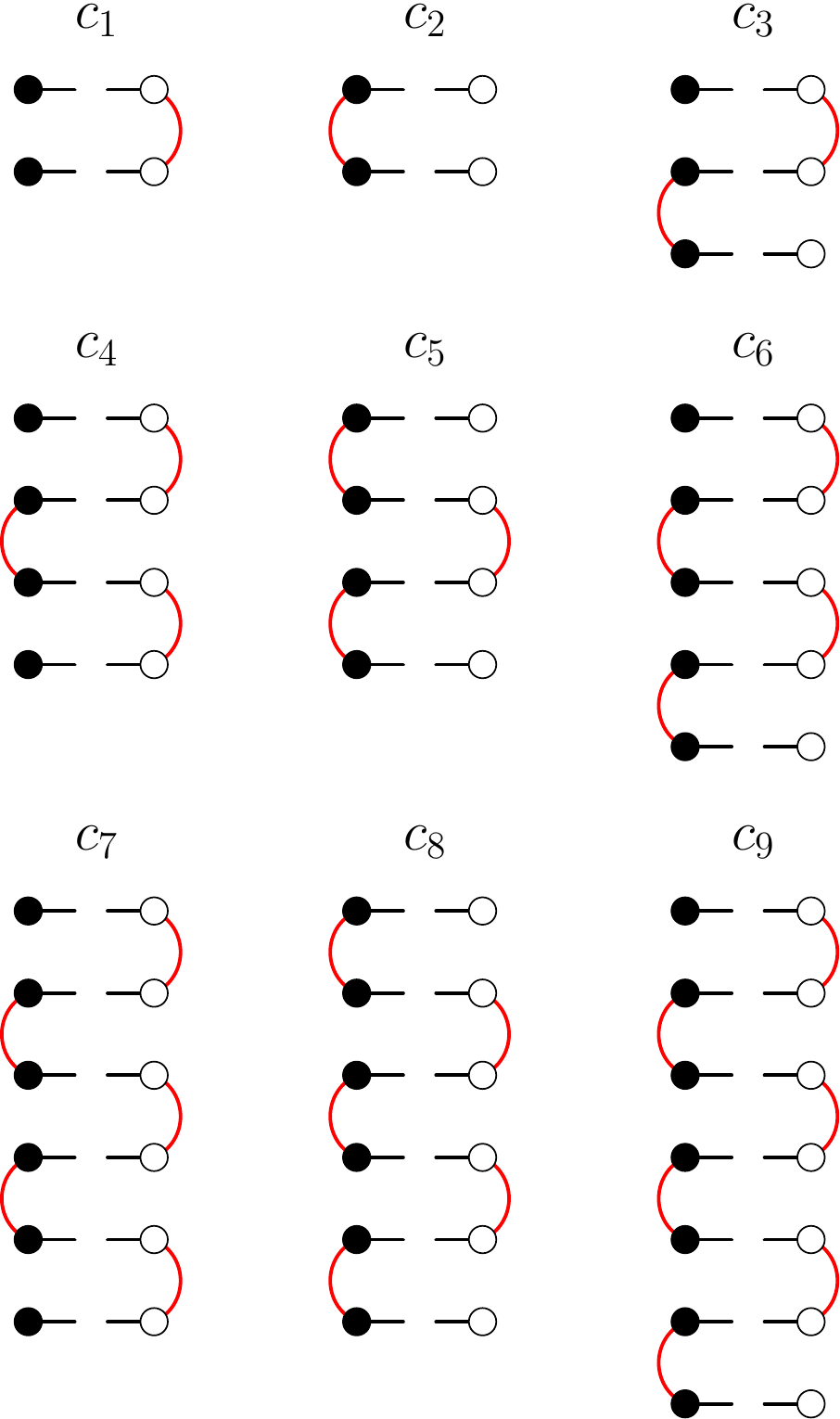}
\caption{\label{fig:chained}
Local chained singlet projection measurements. With each label we associated 
a probability of projecting the multicopy system onto 
singlets (red curves), i.e., HOM anticoalescence. This notation is 
used for expressing both Mahklin's and Jing's invariants in terms 
of multicopy singlet projections. Note that every low rank composite 
singlet projection can be observed as an invent in a more complex 
interferometer designed to measure a higher rank chained singlet projection.}
\end{figure}

\begin{figure}
\includegraphics[width=0.6\linewidth]{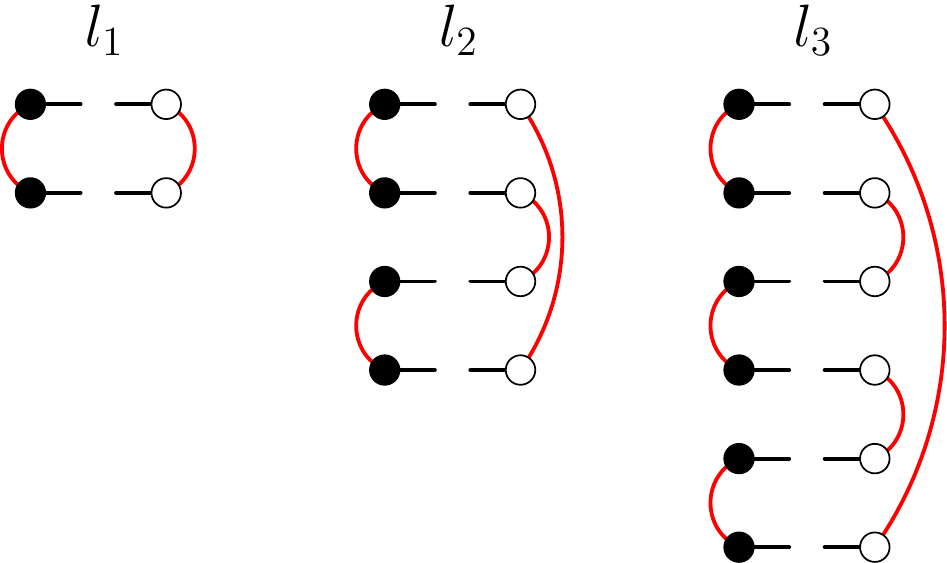}
\caption{\label{fig:looped} The same as in Fig.~\ref{fig:chained} 
but for looped singlet measurement sequences. 
Note that all measurement corresponding to diagrams from $c_1$ to $c_8$ 
can be implemented by an interferometer
designed for measuring one of the looped diagrams $l_n$ for $n=1,2,3$.}
\end{figure}

\begin{figure}
\includegraphics[width=0.6\linewidth]{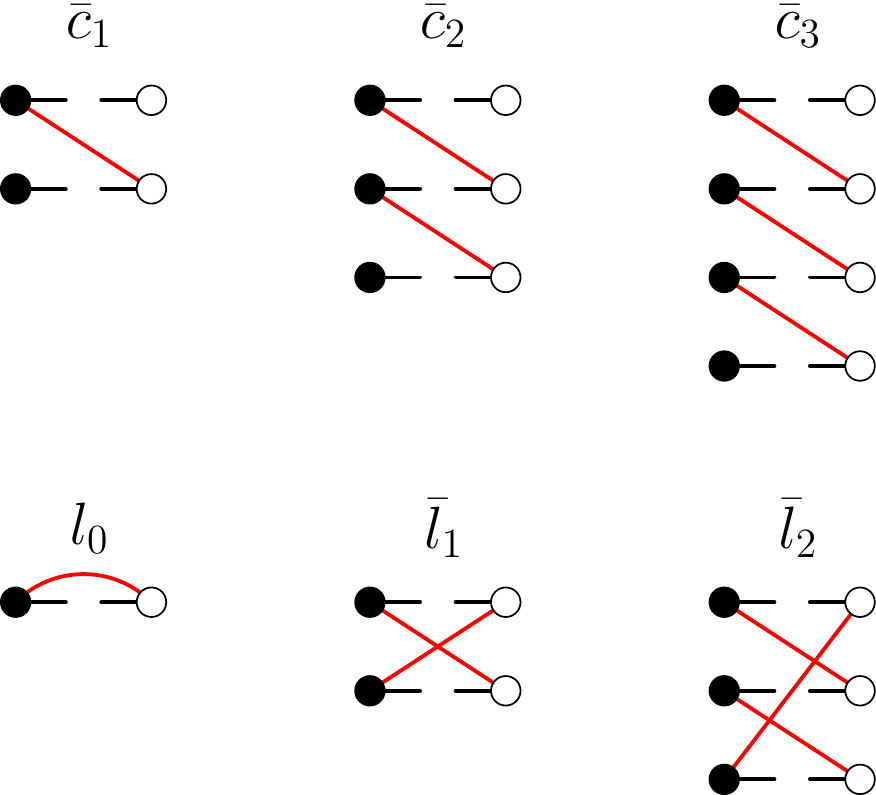}
\caption{\label{fig:nonlocal} Nonlocal chained and looped singlet projection operations. 
With each label we associated a probability of projecting the multicopy system onto 
singlets (red curves). This notation is used for expressing both Mahklin's invariants 
in terms of multicopy singlet projections. Note that measurements
$\bar c_1$ and $\bar c_2$ can can be performed in interferometers
designed for measuring $\bar l_1$ (only $\bar c_1$) or $\bar l_2$.}
\end{figure}

It can be shown by direct calculations that a singlet projection 
\begin{equation}\label{eq:Pproj}
\hat P_{k,l}^-=\tfrac{1}{4}(2\hat\sigma_0\otimes\hat\sigma_0-\hat\sigma_i\otimes\hat\sigma_i)_{a,b}
=(|\Psi^-\rangle\langle\Psi^-|)_{k,l}
\end{equation}
performed on photons in modes $k$ and $l$ is equivalent to a two-particle observable 
$(\hat\sigma_i\otimes\hat\sigma_i )_{k,l}=\hat\sigma_0^{\otimes 2}-4 \hat P^-_{k,l}$ for particles 
$k$ and $l$. 
This projection on a singlet state is graphically represented
throughout this paper as a red curve 
(see Figs. \ref{fig:chained}-\ref{fig:nonlocal}). 
The multiple copies of a  given two-qubit state are depicted as  dashed lines
with black and white ends representing the potentially
quantum-correlated subsystems of Alice and Bob. 

When we have access to multiple copies of the same 
bipartite system we can perform singlet projections between various qubits.
However, not every possible sequence of projections is needed for determining the 
values of local invariants. These sequences depend on the particular invariants.
For the invariants discussed in this paper  we can group the possible sets of projections into
local chained projections (singlet projections are performed only locally, see Fig.~\ref{fig:chained}),
local looped projections (similar as chained projections, but all qubits are
paired,  see Fig.~\ref{fig:looped}), and nonlocal projections 
(singled projections made on systems that are locally separated , see Fig.~\ref{fig:nonlocal}).  
We will demonstrate that,  if we analyze only $J$
invariants~\cite{Jing2015} we do not need to apply the nonlocal projections.

Note that some of the projections shown in Figs.~\ref{fig:chained}--\ref{fig:nonlocal}
require a high number of copies of a given two-qubit state and performing
experiments with a large number of photon pairs may be very challenging~\cite{Pan12RMP}.
However, there are experimental works using multiple copies of
a two-qubit state to measure nonlinear properties of the quantum 
system~\cite{Bovino05PRL,Lemr16collect,Bartkiewicz17fef}. 

\subsection{Reducing W-state projection to singlet projections}
The second prime measurement that appears in the most 
complex of Makhlin's invariants is a three-particle
observable
\begin{eqnarray}\label{eq:Wproj}
\hat W_{k,l,m} &=&(|W_0\rangle\langle W_0| + |W_1\rangle\langle W_1|\nonumber\\
&&-|W_2\rangle\langle W_2|-|W_3\rangle\langle W_3|)_{k,l,m}, 
\end{eqnarray}
where
$|W_{0}\rangle_{k,l,m}=(|HHV\rangle + e^{i2\pi/3}|HVH\rangle + e^{i4\pi/3}|VHH\rangle)_{k,l,m}/\sqrt{3}$,
$|W_{1}\rangle_{k,l,m}=(|VVH\rangle + e^{i2\pi/3}|VHV\rangle + e^{i4\pi/3}|HVV\rangle)_{k,l,m}/\sqrt{3}$,
$|W_{2}\rangle_{k,l,m}=(|HHV\rangle + e^{-i2\pi/3}|HVH\rangle + e^{-i4\pi/3}|VHH\rangle)_{k,l,m}/\sqrt{3}$,
$|W_{3}\rangle_{k,l,m}=(|VVH\rangle + e^{-i2\pi/3}|VHV\rangle + e^{-i4\pi/3}|HVV\rangle)_{k,l,m}/\sqrt{3}$,
are  $W$-states that manifest only tripartite entanglement. 
This observable emerges while dealing with determinants of matrices formed
by 3 Bloch vectors describing qubits in modes $k,l,m$, i.e,
$ \hat W_{k,l,m} = e_{r,s,t}(\hat{\sigma}_r\otimes\hat{\sigma}_s\otimes\hat{\sigma}_t)_{k,l,m}.$
It is interesting that this measurements quantifies the imbalance 
between the probabilities of 3-qubit state belonging
to two subspaces (one spanned by $|W_0\rangle_{k,l,m}, |W_1\rangle_{k,l,m}$ and 
the other by $|W_2\rangle_{k,l,m}, |W_3\rangle_{k,l,m}$) 
being complex conjugates of themselves. 
The complex conjugation of a state is associated with time reversal symmetry
and $\hat W_{k,l,m}$ measurements break this symmetry. Thus,
such a measurement can discriminate spins rotating in 
the opposite directions. 
This measurement can be also interpreted as a way
of distinguishing left-handed and right-handed coordinate system
formed by Bloch vectors corresponding to the 3 measured 
qubits. This is a simple example of quantum supremacy,
where a projection on an entangled state provides
an answer to the stated problem (calculating an arbitrary 
3 dimensional determinant) much faster that the classical analysis.
Note that there is an elegant method of projecting a 3-photon state
on a $W$-state~\cite{Tashima08PRA,Tashima09NJP,Tashima10PRL}. However, 
this method works with limited probability and would not allow us 
to distinguish between the pair of states $(|W_0\rangle_{k,l,m}, |W_1\rangle_{k,l,m})$ and 
 $(|W_2\rangle_{k,l,m}, |W_3\rangle_{k,l,m})$.

\begin{figure}
\includegraphics[width=0.95\linewidth]{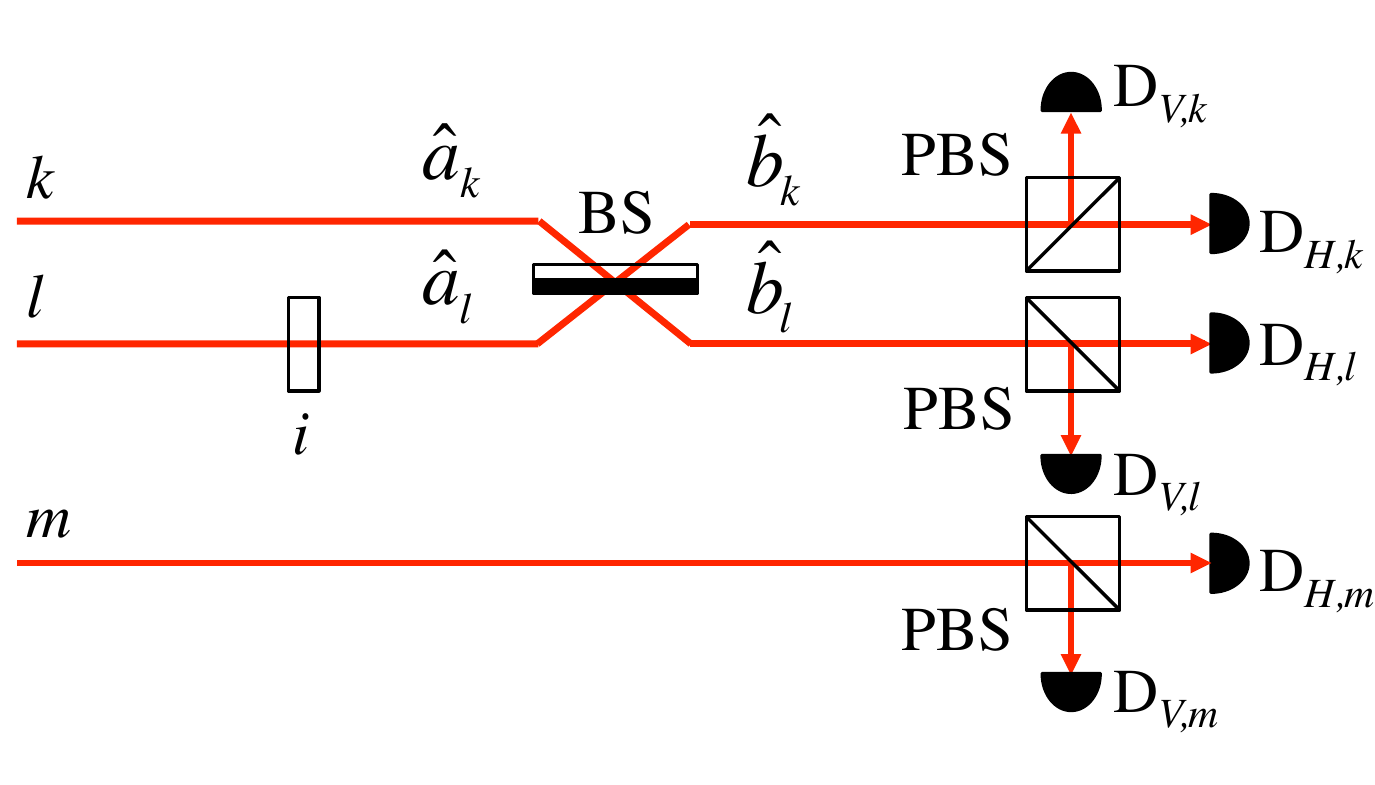}
\caption{\label{fig:circuit} Optical circuit for implementing linear-optical measurement of observable
$\hat w_{k,l,m}$ given in Eq.~(\ref{eq:Wklm}).
The circuit implement is composed of a $\pi/2$ phase shift corresponding to a phase factor $i,$
a balanced  beam splitter (BS) described by Eq.~(\ref{eq:BS}), polarizing beam splitters (PBSs), and detectors 
$\mathrm{D}_{p,n}$ that count photons of polarization $p=H,V$ in spatial modes $n=k,l,m$.
This circuit registers the outcome $+1$ if the following triples of detectors register a photon each, i.e.,  
$(\mathrm{D}_{V,k},\mathrm{D}_{H,k},\mathrm{D}_{V,m}),$ 
$(\mathrm{D}_{V,l},\mathrm{D}_{H,l},\mathrm{D}_{V,m}),$ 
$(\mathrm{D}_{V,k},\mathrm{D}_{H,l},\mathrm{D}_{H,m}),$ 
and$(\mathrm{D}_{H,k},\mathrm{D}_{V,l},\mathrm{D}_{V,m}).$ 
Similarly, it registers $-1$ for the triples 
$(\mathrm{D}_{V,k},\mathrm{D}_{H,k},\mathrm{D}_{H,m}),$
$(\mathrm{D}_{V,l},\mathrm{D}_{H,l},\mathrm{D}_{H,m}),$ 
$(\mathrm{D}_{V,k},\mathrm{D}_{H,l},\mathrm{D}_{V,m}),$ 
and $(\mathrm{D}_{H,k},\mathrm{D}_{V,l},\mathrm{D}_{H,m}).$
The other possible detection events are associated with value $0.$}
\end{figure}

It turns out that we can implement the $W$-state projection by HOM interference by using its
alternative representation, i.e., 
\begin{equation}\label{eq:WprojA}
\hat W_{k,l,m}=\hat w_{k,l,m}+\hat w_{l,m,k}+\hat w_{m,k,l}
\end{equation}
projection with 3 configurations of HOM interferometer using the circuit depicted in Fig.~\ref{fig:circuit}, where
\begin{equation}\label{eq:Wklm}
\hat w_{k,l,m} = \left[2\hat S(|\Psi^-\rangle \langle \Psi^- |-|\Psi^+\rangle \langle \Psi^+ |)\hat S^\dagger\otimes\hat\sigma_z\right]_{k,l,m}
\end{equation}
for photons in spatial modes $k,l,m$ and where $\hat S=\mathrm{diag}[1,1,i,i]$ is a single-qubit phase gate.
This measurement reveals an interesting feature of
quantum physics. By interference we can learn about the
mutual orientation of three real  (Bloch) vectors 
(decide if they are ordered in a way that form left or right-handed system) 
in only three measurements. 
Using a direct approach one has to measure all 3 components of 
all 3 vectors (i.e., 9 measurements in total in the general case of quantum correlated 3-qubit state). 
Thus, in this case we deal with quadratic speed-up.  If one could measure the 
$W$-state projections 
directly, this speed up would be even greater.

\section{Local unitary invariants of two-qubit states}

\subsection{Makhlin's Invariants}
The invariants described by Makhlin in Ref.~\cite{Makhlin2002}
can be expressed in terms of the correlation matrix 
$\hat{\beta}= \mathrm{tr}[(\hat\sigma_i \otimes\hat\sigma_j)\hat\rho]$, and
the Bloch vectors $\mathbf{s}=\mathrm{tr}[({\hat\sigma}_i\otimes{\hat\sigma}_0)\hat\rho]$
and $\mathbf{p}=\mathrm{tr}[(\hat\sigma_0\otimes{\hat\sigma}_j)\hat\rho]$.
The matrices $\hat\sigma_i$ for $i=0,\,1,\,2,\,3$ are the 
Pauli matrices with $\hat\sigma_0$ being the single-qubit identity matrix. 
These invariants~\cite{Makhlin2002} are given as 
$I_1=\det\hat\beta$,
$I_2=\mathrm{tr}(\hat\beta^T\hat\beta)$,
$I_3=\mathrm{tr}(\hat\beta^T\hat\beta)^2$, 
$I_4=\mathbf{s}^2$,
$I_5=[\mathbf{s}\hat\beta]^2$, 
$I_6 = [\s\hat\beta\hat\beta^T]^2$, 
$I_7 =\mathbf{p}^2$, 
$I_8 =[\hat\beta\mathbf{p}]^2$, 
$I_9=[\hat\beta^T\hat\beta\p]^2$,
$I_{10}=(\s,\;\s\hat\beta\hat\beta^T,\;\s[\hat\beta\hat\beta^T]^2)$,
$I_{11}=(\p,\;\hat\beta^T\hat\beta\p,\;[\hat\beta^T\hat\beta]^2\p)$,
$I_{12} =\mathbf{s}\hat\beta\mathbf{p}$, 
$I_{13}=\s \hat\beta\hat\beta^T\hat\beta \p$,
$I_{14} = e_{ijk}e_{lmn}s_ip_l\beta_{jm}\beta_{kn}$, 
$I_{15}=(\s,\;\s\hat\beta\hat\beta^T,\;\hat\beta\p)$,
$I_{16}=(\s\hat\beta,\;\p,\;\hat\beta^T\hat\beta\p)$,
$I_{17}=(\s\hat\beta,\;\s\hat\beta\hat\beta^T\hat\beta,\;\p)$,
$I_{18}=(\s,\;\hat\beta\p,\;\hat\beta\hat\beta^T\hat\beta\p)$.
Here $({\bf a},{\bf b},{\bf c})$ stands for the triple scalar product ${\bf
a}\cdot({\bf b}\times{\bf c})$ and $e_{ijk}$ is the Levi-Civita symbol.

As shown is Ref.~\cite{Bartkiewicz17hom}, the Makhlin's invariants relevant to measuring
entanglement in terms of  negativity are as follows
 \begin{eqnarray}
I_1 &=&  -\tfrac{8}{3}\lbrace l_{0} [l_{0} (4l_{0} -3)+6 (\bar{c}_{1}-2 \bar{l}_{1})] \nonumber\\
&& +3 \bar{l}_{1} -6\bar{c}_{2}+8\bar{l}_{2}\rbrace,\nonumber\\
I_2 &=&  1 + 16 {l}_{1} - 4 ({c}_{2} + {c}_{1}),\nonumber\\
I_3 &=&  1+256 \left({c}_{2}^2+4 {c}_{3}+{c}_{1}^2+{l}_{2}\right)-8 ({c}_{2}+{c}_{1})\nonumber\\
I_4 &=&  1-4{c}_{2}, \label{eq:MinvG}\\
I_5 &=&  -4 {c}_{1} +32 {c}_{3} - 64 {c}_{5}+(1- 4 {c}_{2})^2,\nonumber\\
I_7 &=&  1-4{c}_{1},\nonumber\\
I_8 &=&  -4 {c}_{2} + 32 {c}_{3} - 64 {c}_{4} + (1 - 4 {c}_{1})^2,\nonumber\\
I_{12} &=& 1 + 16 {c}_{3} - 4 ({c}_{2} + {c}_{1}),\nonumber\\
I_{14} &=&  16 [l_{0}^2 (1 - 4 \bar{c}_{1})  +2 l_{0} (  4 \bar{c}_{2}-\bar{c}_{1}) \nonumber\\
&&  - \bar{l}_{1}  + 4 \bar{c}_{1} \bar{l}_{1}  + 2 \bar{c}_{2} - 8 \bar{c}_{3} ],\nonumber
\end{eqnarray}
where the relevant 13 terms  
$l_{0},$ ${c}_{2},$ $\bar{c}_{1},$ ${c}_{1},$
${l}_{1},$ ${c}_{3},$ $\bar{l}_{1}$ $\bar{c}_{2},$
$\bar{l}_{2},$  ${c}_{5},$  ${c}_{4},$ 
${l}_{2},$  $\bar{c}_{3},$ 
are singlet projections depicted in Figs.~\ref{fig:chained}--\ref{fig:nonlocal} and can be measured utilizing only 
HOM interfernce.
Similarly, we find the following 3 of the remaining  invariants 
 \begin{eqnarray}
I_6 &=& 1 - 1024 {c}_{8} - 4 (3 {c}_{2} + 2 {c}_{1}) \nonumber\\
&&+  16 (3 {c}_{2}^2 + 4 {c}_{3} + 2 {c}_{2} {c}_{1} + {c}_{1}^2) \nonumber\\ 
&&- 64 ({c}_{2}^3 + 4 {c}_{2} {c}_{3} + 2 {c}_{5} + 2 {c}_{3} {c}_{1}+ {c}_{4})\nonumber\\
&&  + 256 ({c}_{3}^2 + 2 {c}_{2} {c}_{5} +2 {c}_{6} )\nonumber\\
I_9 &=&  1-   1024 {c}_{7}- 4 (2 {c}_{2} + 3 {c}_{1})  \nonumber\\
&&+16 ({c}_{2}^2 + 4 {c}_{3} + 2 {c}_{2} {c}_{1} + 3 {c}_{1}^2)  \nonumber\\
&& - 64 (2 {c}_{2} {c}_{3} + {c}_{5} + 4 {c}_{3} {c}_{1} + {c}_{1}^3 + 2 {c}_{4}) \\
&&  +  256 ({c}_{3}^2 +2 {c}_{6} + 2 {c}_{1} {c}_{4}),\nonumber\\
I_{13} &=& 1 + 256 {c}_{6} - 8 ({c}_{2} + {c}_{1})  \nonumber\\
&& +16 ({c}_{2}^2 + 3 {c}_{3} + {c}_{2} {c}_{1} + {c}_{1}^2) \nonumber\\
&& -64 [{c}_{5} + {c}_{3} ({c}_{2} + {c}_{1}) + {c}_{4}],
 \nonumber
\end{eqnarray}
that can be measured using very similar interferometers 
composite HOM to these described in Ref.~\cite{Bartkiewicz17hom}.
These can be designed as explained in Ref.~\cite{Bartkiewicz17hom}, i.e.,
by constructing interferometers that would at best (if all the detector 
pairs detect anticoalescence) measure the values of $c_7$ or $c_8$, and 
for other combinations of aniticoalescence  and coalescence 
events would measure polynomials of $c_n$ for $n=1,2,3,4,5,6$.
The remaining six remaining invariants $I_n$ for $n=10,11,15,16,17,18$
require a new approach. These six  invariants are needed only to bound the signs 
of the components of the $\mathbf{s}$ and $\mathbf{p}$
vectors. Thus, their absolute values are not important. 
We cannot measure them directly only with HOM interference
limited only to coalescence  and anticoalescence detection. 
This is because in order to estimate the value of 
a three-particle observable 
$\langle e_{ijk}\hat{\sigma}_i\otimes\hat{\sigma}_j\otimes\hat{\sigma}_k \rangle_{k,l,m}$
for particles $k,$ $l,$ and $m$ one needs to measure $\hat W_{k,l,m}$ defined in Eq.~(\ref{eq:Wproj}). 
The physical interpretation of this $\hat W$ measurement 
is the difference of the probabilities of Bloch vectors of the three qubits forming left-hand and
right-hand coordinate system.
If one performs only the anticoalescence  detection, at best one measures
$\hat W^2_{k,l,m}=\sum_{n=0,1,2,3} |W_n\rangle\langle W_n|_{k,l,m}=P_{k,l}+P_{k,m}+P_{l,m},$
which  does not break the time reversal symmetry. 
Hence,  only with simple 
interferometers we can measure only $I_n^2$ for 
$n=10,11,15,16,17,18,$
where the sign is lost and it makes the invariants useless.
However, $\hat W$ measurement can be performed indirectly
as explained by Eq.~(\ref{eq:Wklm}) and in Fig.~\ref{fig:circuit}.

It is very interesting to observe that one would need three-particle
measurements to measure directly some of the local two-particle invariants. 
However, these handness invariants are special as they reveal mutual orientation 
of the Bloch vector components
of the subsystems of density matrix $\hat \rho$ (i.e., the signs of $s_i,p_i$ for $i=1,2,3$)
\cite{Makhlin2002}, while other $I$ invariants could be used only to determine the absolute values.
Thus, any locally invariant properties of a two-qubit state can be assessed by using only 
singlet and $W$-state projections on multiple copies of the two-qubit system. The latter can be expressed by 
modified $|\Psi^\pm\rangle$  projections and $\hat\sigma_z$ measurements as shown in Eq~(\ref{eq:Wklm}).
Hence, we can measure the $J$ invariants with only HOM interference and $\hat\sigma_z$ measurement.
The exact experimental procedure for measuring invariants $I_n$ for 
$n=10,11,15,16,17,18$ is straightforward, but it would take much space to cover in detail.
For sake of clarity of the paper we list only the partial observations needed for
such measurements within the above-described framework in  Appendix~\ref{sec:appendix}.
All these observations are local.

\subsection{Jing's et al. invariants}
It turns out that we do not need $W$ state measurement or $\hat \sigma_z$ to 
check if a two two-qubit states are equivalent up to local unitaries.
Remarkably it was shown by Jing et al.~\cite{Jing2015} that 
there are 12 local invariants that are equivalent
to the set of 18 Makhlin's invariants. This means that both 
sets of invariants are sufficient to decide if any pair of two-qubit states
is locally equivalent.
The 6 Makhlin's invariants $I_n$ for $n=10,11,15,16,17,18$
are inequivalent to  trivial polynomials of Jing's invariants, as 
at the most fundamental level they cannot be reduced to simple singlet 
projections on multiple copies. However, we find that Jing's invariants 
can be related to other Makhlin's invariants via singlet projections
in the following way
 \begin{eqnarray}
J_{1} &=& \mathrm{tr}(\hat\beta^T\hat\beta)=I_2,\nonumber\\
J_{2} &=& \mathrm{tr}(\hat\beta^T\hat\beta)^2=I_3,\nonumber\\
J_{3} &=& \mathrm{tr}(\hat\beta^T\hat\beta)^3=\tfrac{1}{2}(6I^2_1-I_2^3+ 3 I_2 I_3),\nonumber\\
J_{4} &=& \mathbf{s}^2=I_4,\nonumber\\
J_{5} &=& [\mathbf{s}\hat\beta]^2=I_5,\nonumber\\
J_{6} &=& [\mathbf{s}\hat\beta\hat\beta^T]^2=I_6,\nonumber\\
J_{7} &=& \mathbf{p}^2=I_7,\\ 
J_{8} &=&  [\hat\beta\mathbf{p}]^2=I_8,\nonumber\\ 
J_{9} &=& [\hat\beta^T\hat\beta\mathbf{p}]^2=I_9,\nonumber\\ 
J_{10} &=& \mathbf{s}\hat\beta\mathbf{p} =I_{12},\nonumber\\
J_{11} &=& \mathbf{s}\hat\beta\hat\beta^T\hat\beta\mathbf{p} =I_{13},\nonumber\\
J_{12} &=& \mathbf{s}\hat\beta\hat\beta^T\hat\beta\hat\beta^T\hat\beta\mathbf{p} .\nonumber
\end{eqnarray}
In particular we can also express $J_3$ and $J_{12}$ as
$J_3=4096 l_3 - 3072(c_7+c_8)+768(2c_6+c_4c_1+c_5c_2+c_3^2)
-64[3c_4 +3c_5 +6c_3(c_1+c_2)+c_1^3+c_2^3] 
+ 48(2c_3+c_1^2 +c_2^2+c_1c_2)-12(c_1+c_2)+1$ and
$J_{12}=4096c_9 - 1024(c_7+c_8+c_6c_1+c_6c_2+c_3c_4+c_3c_5)
+768(3c_6+c_4c_1+c_5c_2+c_3^2)
-64[2c_4 +2c_5 +6c_3(c_1+c_2)+c_1^3+c_2^3 +c_1^2 c_2 +c_2^2c_1]
+ 16(5c_3+3c_1^2 +3c_2^2+4c_1c_2)-12(c_1+c_2)+1,$ respectively.
The interferometer for measuring $J_{12}$ is equivalent to
an interferometer designed for measuring $c_9$ in the case of detecting 
only anticoalescene events  and other polynomials
of $c_n$ for $n=1,2,3,4,5,6,7,8$ for specific combinations
of anticoalescene and coalescence events in the relevant detector pairs.
Now we can make two interesting observations. 
Firstly, unlike the Makhlin's invariants 
all the invariants can be expressed by only local loops and chains. Secondly,
note that the measurement of $I_1$ includes only nonlocal singlet projections
which are fundamentally different from those which measure  $I_1^2$ (only local singlet projections).
This can be seen by expressing $I^2_1$ only by $J_3,$ $I_2,$ 
and $I_3,$ which all three can be measured using only local singlet projections.
The operational simplicity of $J$ invaraints has its price, but also some benefits. 
For example, due to the lost information about the sign of $I_1$ and no apparent 
way of extracting the value of $I_{14}$, we cannot calculate a value of negativity 
using solely $J$ invariants. On the other hand, if one uses $J$ invariants there 
is no need for performing $\hat W$ measurement to check,  if two states 
are equivalent up to local unitaries. Moreover, all the projections needed here are local.
Hence, despite its benefits for verifying local equivalence of states, Jing's
invariants appear to not be useful for measuring quantum entanglement 
(i.e., for measuring negativity we need both $I_1$ and $I_{14}$ invariants 
that are measured via nonlocal measurements). However, still they can be applied 
to measuring nonlocality, as we demonstrate in the following section.

\section{Two invariant-based methods for measuring Bell-CHSH nonlocality}

\begin{figure}
\includegraphics[width=0.9\linewidth]{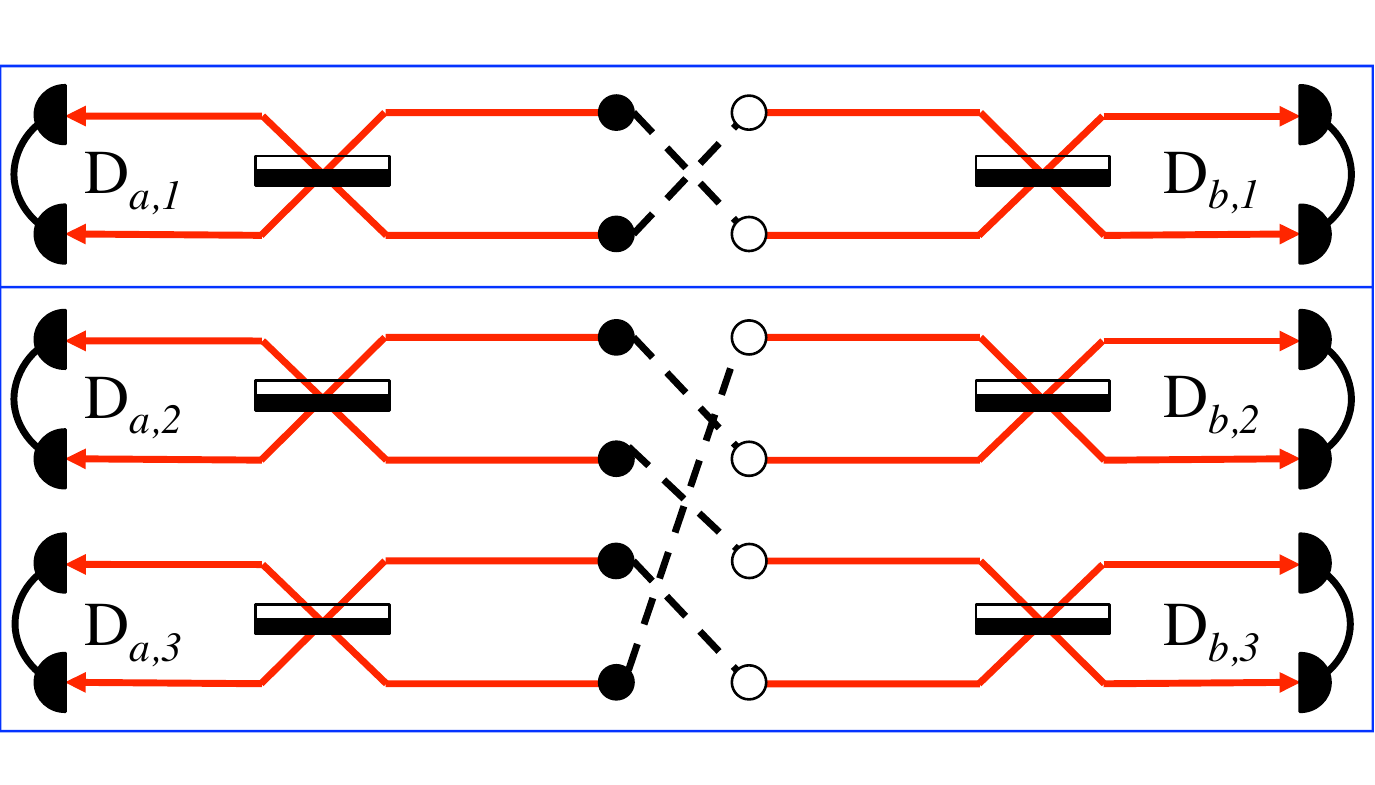}
\caption{\label{fig:I1} 
Interferometric configurations for measuring two independent  
invariants $I_2$ and $I_3$ (or $J_1$ and $J_2$)
associated with looped singlet projections $l_1$ and $l_2$
with two and four copies of polarization-encoded $\hat\rho$. 
Subsystems of a single copy
are depicted as black and white discs connected with dashed lines.
Photons interfere on beam splitters BS and their
coalescence or anti-coalescence is detected by detector modules D$_{a,n}$
(Alice) and D$_{b,n}$ (Bob) for $n=1,2,3$. 
For detailed analysis of all the possible detection events
see Tab.~\ref{tab:1}.
To measure these invariant one needs to access at most 2 or 4 copies of the investigated state for $I_2$ and $I_3$ (or $J_1$ and $J_2$),
respectively. All the measurements are local.}
\end{figure}

\begin{figure}
\includegraphics[width=0.9\linewidth]{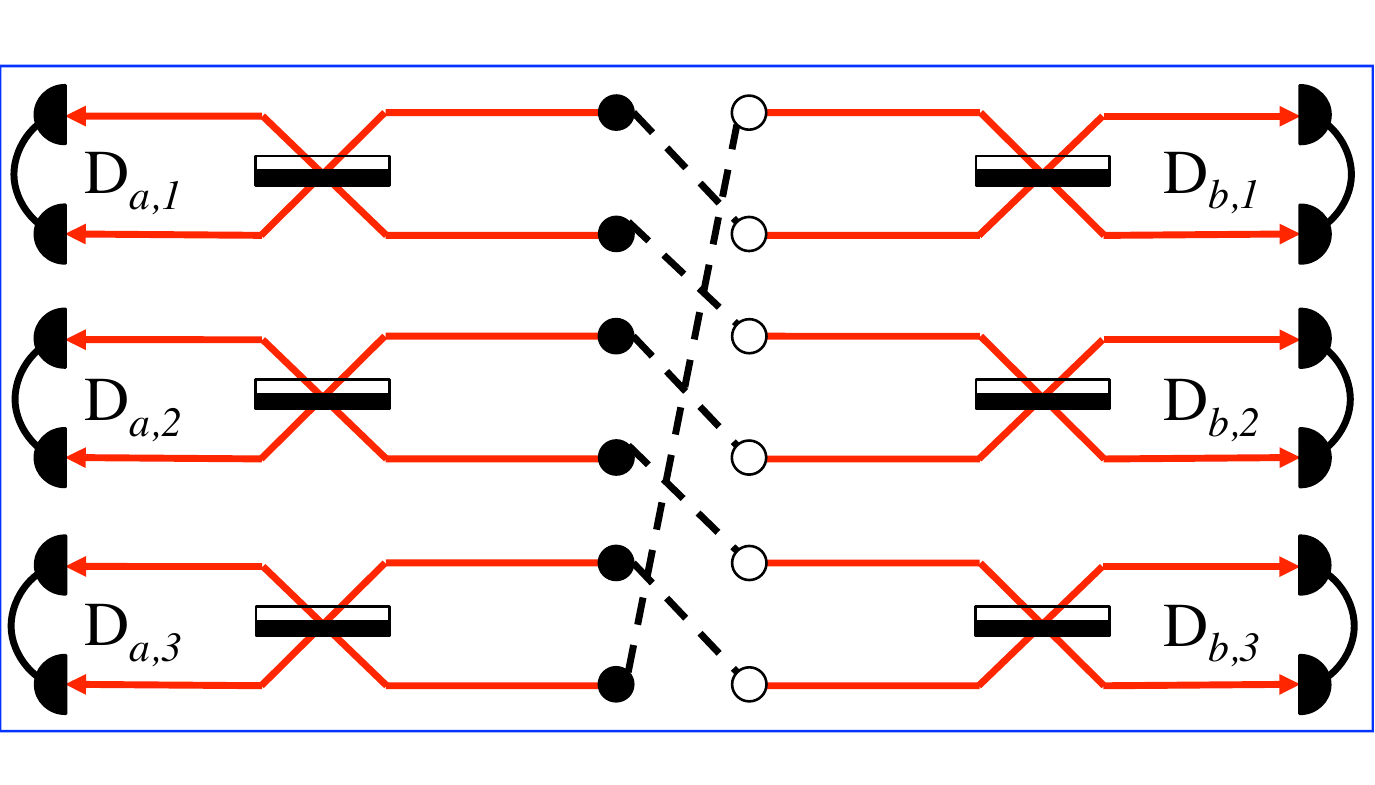}
\caption{\label{fig:I2}  Same as in Fig.~\ref{fig:I1}, but for
an interferometer measuring $J_3$. To measure this invariant one needs to access 6 copies of the investigated state.
For detailed analysis of all the possible detection events
see Tab.~\ref{tab:2}.}
\end{figure}

\begin{figure}
\includegraphics[width=0.9\linewidth]{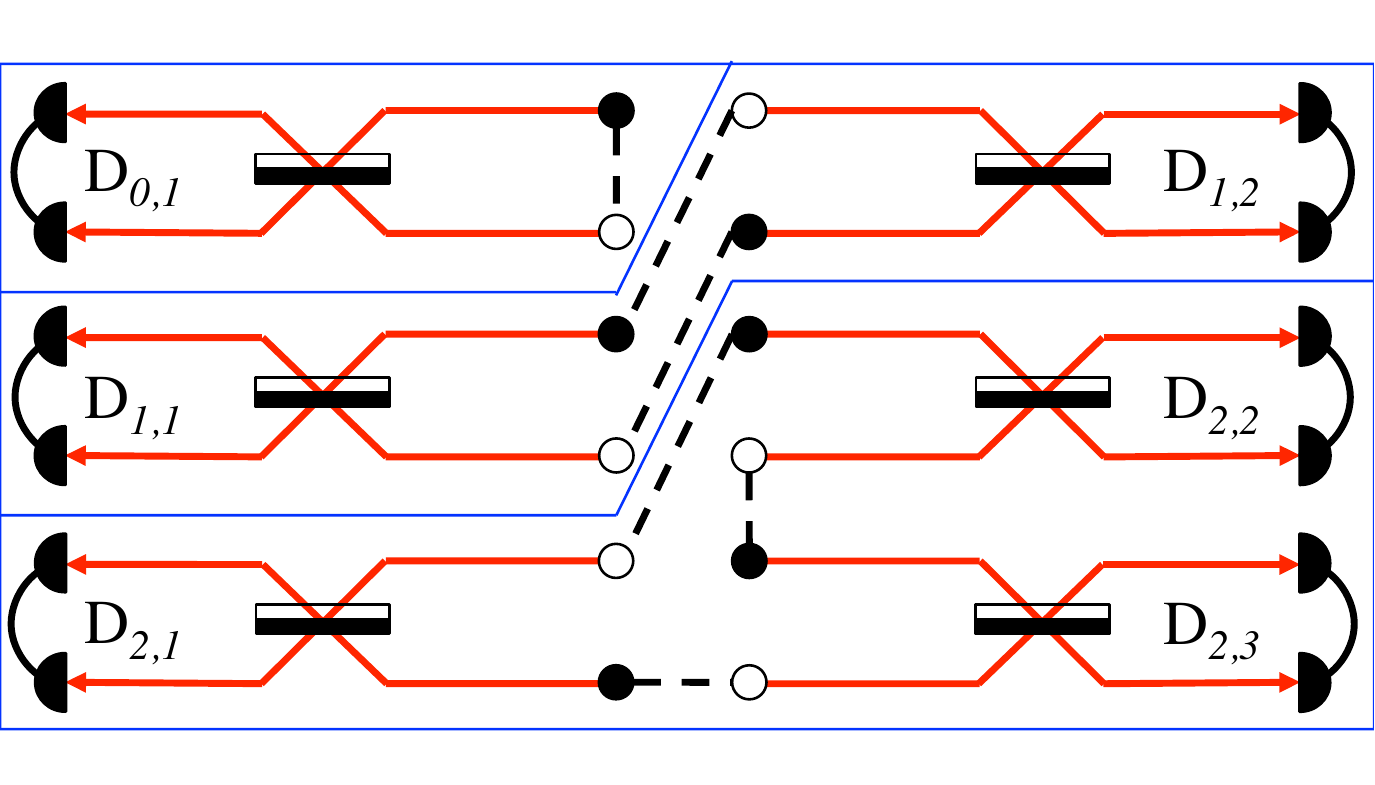}
\caption{\label{fig:I3}  Same as in Fig.~\ref{fig:I1}, but for
an interferometer measuring $I_1$  in terms of (from top) 
$l_0,$ $\bar l_1,$ and $\bar l_2$. All the measurements are nonlocal (singlet projections
are performed between subsystems of Alice and Bob).
To measure this invariant one needs to access at most 3 copies of the investigated state. 
For detailed analysis of all the possible detection events
see Tab.~\ref{tab:3}.}
\end{figure}

\begin{table}
\caption{\label{tab:1} 
Interpretation of detection events of the interferometers shown in Fig.~\ref{fig:I1}.
Each couple of detectors  D$_{a,n}$ and $D_{b,n}$ for $n=1,2,3$  detects coalesce or
anti-coalescence for a pair of impinging photons.
The accumulated counts of (anti-) coalescence events can be 
grouped into $c$ coalescence or $s=a+c$ sum of coalescence ($c$)
and anti-coalescence ($a$). The total number of all detection events in D$_{a,1}$ and $D_{b,1}$
is $Z_1$. The remaining detection events accumulate to $Z_2$.
Depending on the measured quantity, one can choose
the required detection events in accord with Fig.~\ref{fig:chained} or Fig.~\ref{fig:looped}.}
\begin{ruledtabular}
\begin{tabular}{ccccccc}
D$_{a,1}$&	D$_{a,2}$&	D$_{a,3}$&	D$_{b,1}$&	D$_{b,2}$&	D$_{b,3}$&	Fig.~\ref{fig:I1}\\
\hline
s&-&-&s&-&-&$Z_1$\\
s&-&-&a&-&-&$Z_1c_1$\\
a&-&	-&	s&-&	-&$Z_1c_2$\\
a&-&	-&	a&-&	-&$Z_1l_1$\\
-&s&s&	-&	s&s&$Z_2$\\
-&s&s&-&s&a&$Z_2c_1$\\
-&s&s&-&a&s&$Z_2c_1$\\
-&s&s&-&a&a&$Z_2c_1^2$\\
-&s&a&-&s&s&$Z_2c_2$\\
-&s&a&-&s&a&$Z_2c_3$\\
-&s&a&-&a&s&$Z_2c_3$\\
-&s&a&-&a&a&$Z_2c_4$\\
-&a&s&-&s&s&$Z_2c_2$\\
-&a&s&-&s&a&$Z_2c_3$\\
-&a&s&-&a&s&$Z_2c_3$\\
-&a&s&-&a&a&$Z_2c_4$\\
-&a&a&-&s&s&$Z_2c_2^2$\\
-&a&a&-&s&a&	$Z_2c_5$\\
-&a&a&-&a&s&	$Z_2c_5$\\
-&a&a&-&a&a&$Z_2l_2$
\end{tabular}
\end{ruledtabular}
\end{table}

\begin{table*}
\caption{\label{tab:2} 
The same as in Tab.~\ref{tab:1}, but for an interferometer shown in Fig.~\ref{fig:I2}}
\begin{ruledtabular}
\begin{tabular}{cccccccccccccc}
D$_{a,1}$&	D$_{a,2}$&	D$_{a,3}$&	D$_{b,1}$&	D$_{b,2}$&	D$_{b,3}$&	Fig.~\ref{fig:I2} &D$_{a,1}$&	D$_{a,2}$&	D$_{a,3}$&	D$_{b,1}$&	D$_{b,2}$&	D$_{b,3}$&	Fig.~\ref{fig:I2}\\
\hline
s&s&s&s&s&s&$Z$&a&s&s&s&s&s&$Zc_2$\\
s&s&s&s&s&a&$Zc_1$&a&s&s&s&s&a&$Zc_2c_1$\\
s&s&s&s&a&s&$Zc_1$&a&s&s&s&a&s&$Zc_2c_1$\\
s&s&s&s&a&a&$Zc_1^2$&a&s&s&s&a&a&$Zc_2c_1^2$\\
s&s&s&a&s&s&$Zc_1$&a&s&s&a&s&s&$Zc_3$\\
s&s&s&a&s&a&$Zc_1^2$&a&s&s&a&s&a&$Zc_1c_3$\\
s&s&s&a&a&s&$Zc_1^2$&a&s&s&a&a&s&$Zc_4$\\
s&s&s&a&a&a&$Zc_1^3$&a&s&s&a&a&a&$Zc_1c_4$\\
s&s&a&s&s&s&$Zc_2$&a&s&a&s&s&s&$Zc_2^2$\\
s&s&a&s&s&a&$Zc_3$&a&s&a&s&s&a&$Zc_2c_3$\\
s&s&a&s&a&s&$Zc_3$&a&s&a&s&a&s&$Zc_2c_3$\\
s&s&a&s&a&a&$Zc_4$&a&s&a&s&a&a&$Zc_3^2$\\
s&s&a&a&s&s&$Zc_1c_2$&a&s&a&a&s&s&$Zc_2c_3$\\
s&s&a&a&s&a&$Zc_1c_3$&a&s&a&a&s&a&$Zc_6$\\
s&s&a&a&a&s&$Zc_1c_3$&a&s&a&a&a&s&$Zc_3^2$\\
s&s&a&a&a&a&$Zc_1c_4$&a&s&a&a&a&a&$Zc_3c_4$\\
s&a&s&s&s&s&$Zc_2$&a&a&s&s&s&s&$Zc_2^2$\\
s&a&s&s&s&a&$Zc_1c_2$&a&a&s&s&s&a&$Zc_2c_3$\\
s&a&s&s&a&s&$Zc_3$&a&a&s&s&a&s&$Zc_5$\\
s&a&s&s&a&a&$Zc_1c_3$&a&a&s&s&a&a&$Zc_6$\\
s&a&s&a&s&s&$Zc_3$&a&a&s&a&s&s&$Zc_2c_3$\\
s&a&s&a&s&a&$Zc_1c_3$&a&a&s&a&s&a&$Zc_3^2$\\
s&a&s&a&a&s&$Zc_4$&a&a&s&a&a&s&$Zc_6$\\
s&a&s&a&a&a&$Zc_1c_4$&a&a&s&a&a&a&$Zc_7$\\
s&a&a&s&s&s&$Zc_2^2$&a&a&a&s&s&s&$Zc_2^3$\\
s&a&a&s&s&a&$Zc_5$&a&a&a&s&s&a&$Zc_2c_5$\\
s&a&a&s&a&s&$Zc_2c_3$&a&a&a&s&a&s&$Zc_2c_5$\\
s&a&a&s&a&a&$Zc_6$&a&a&a&s&a&a&$Zc_8$\\
s&a&a&a&s&s&$Zc_2c_3$&a&a&a&a&s&s&$Zc_2c_5$\\
s&a&a&a&s&a&$Zc_3^2$&a&a&a&a&s&a&$Zc_8$\\
s&a&a&a&a&s&$Zc_1c_4$&a&a&a&a&a&s&$Zc_8$\\
s&a&a&a&a&a& $Zc_7$ &a&a&a&a&a&a&$Zl_3$\\
\end{tabular}
\end{ruledtabular}
\end{table*}

\begin{table}
\caption{\label{tab:3} 
Same as in Tab.~\ref{tab:1}, but for interferometers measuring nonlocal singlet projections form Fig.~\ref{fig:nonlocal}. 
Each couple of detectors  D$_{0,1},$ D$_{1,1},$ D$_{1,2},$ and 
D$_{2,n},$ for $n=1,2,3$  detects coalesce or anti-coalescence for a pair of impinging photons.}
\begin{ruledtabular}
\begin{tabular}{ccccccc}
D$_{0,1}$&	D$_{1,1}$&	D$_{1,2}$&	D$_{2,1}$&	D$_{2,2}$&	D$_{2,3}$&	Fig.~\ref{fig:I3}\\
\hline
s&-&-&-&-&-&$Z_1$\\
a&-&-&-&-&-&$Z_1l_0$\\
-&s&s&-&-&	-&$Z_2$\\
-&s&a&	-&-&	-&$Z_2\bar c_1$\\
-&a&s&	-&	-&-&$Z_2\bar c_1$\\
-&a&a&-&-&-&$Z_2\bar l_1$\\
-&-&-&s&s&s&$Z_3$\\
-&-&-&s&s&a&$Z_3\bar c_1$\\
-&-&-&s&a&s&$Z_3\bar c_1$\\
-&-&-&s&a&a&$Z_3\bar c_2$\\
-&-&-&a&s&s&$Z_3\bar c_1$\\
-&-&-&a&s&a&$Z_3\bar c_2$\\
-&-&-&a&a&s&$Z_3\bar c_2$\\
-&-&-&a&a&a&$Z_3\bar l_2$
\end{tabular}
\end{ruledtabular}
\end{table}

Nonclassical correlations of polarizations can be measured by measuring
only the eigenvalues of $\hat R=\hat\beta\hat\beta^T$ matrix. As it was experimentally
demonstrated in Ref.~\cite{Bartkiewicz17fef}, if one works with
two copies of a density matrix, only six measurements are required
to learn the eigenvalues $r_n$ for $n=1,2,3$. These eigenvalues can be used to express
not only the maximal degree of Bell-CHSH inequality violation but also, e.g., fully-entangled fraction,
and entropic entanglement witness for symmetric states~\cite{Bartkiewicz17fef}.
The Horodecki measure of Bell (or CHSH) nonlocality 
can be expressed as~\cite{Horodecki96PLA}: 
\begin{equation}\label{eq:M}
M=\mathrm{Tr} \hat R - \min[\mathrm{eig}(\hat R)]-1.
\end{equation}
Its values are positive if the Bell-CHSH inequality 
is violated and it reaches the maximum $M=1$ 
for maximally-entangled states.
To express the
fully-entangled fraction
\begin{equation}\label{eq:X}
f=\tfrac{1}{4}\left(\mathrm{Tr}\sqrt{\hat R} +1\right),
\end{equation}
which can be used to quantify the fidelity of many
entanglement-based protocols
\cite{Horodecki96PRA,Badziag00PRA,Grondalski02PLA,Bartkiewicz17fef},
one needs to calculate the square roots of the eigenvalues.
Finally, the sum of eigenvalues of $\hat R$ can be  used directly to express
the entropic entanglement witness $E$ for equal purities of subsystems $a$ and $b$ (i.e.,
$\mathrm{Tr}\rho^2_a=\mathrm{Tr}\rho^2_b$), and it reads as 
\begin{eqnarray}
E=\mathrm{2(Tr}\hat\rho_{a,b}^2
- \min[\mathrm{Tr}\hat\rho^2_a,\mathrm{Tr}\hat\rho^2_b])
=\tfrac{1}{2}(\mathrm{Tr}\hat R-1)\label{eq:E}.
\end{eqnarray}
The measured value of this witness is positive, if it detects quantum entanglement and is negative otherwise.
The spectrum of $\hat R$ can be calculated by measuring the first three  invariants
of Jing by applying only local projections  
or by measuring the first three invariants of Makhlin on fewer copies of the investigated
state, but with using nonlocal projections. 
The spectrum of a three-dimensional matrix is given by the roots of the following 
polynomial in $r$ in terms of $J$-invariants
\begin{equation}
-r^3+J_1 r^2 +(J^2_1-J_2)r+J^3_1+2J_3-3J_1J_2=0
\end{equation}
or in terms of $I$-invariants
\begin{equation}
-r^3+I_2 r^2 +(I^2_2-I_3)r+I^3_2+(6I^2_1-I_2^3)=0.
\end{equation}
A similar approach can be used for determining eigenvalues of density matrices~\cite{Tanaka14PRA}.
The specialized interferometers  designed for measuring these projections
are depicted in Figs.~\ref{fig:I1},\ref{fig:I2} and~\ref{fig:I1},\ref{fig:I3} for invariants $J$ and $I$, respectively.
The  sets of projections necessary to determine the 
eigenvalues $r$ in the case of working with Makhlin's and Jing's et al. invariants
are listed in Tabs.~\ref{tab:1},\ref{tab:2}, and \ref{tab:3}.
From these sets of projections it is apparent that we can learn the value of, e.g., optimal CHSH
nonlocality by using fewer copies of $\hat \rho$ in the case of nonlocal HOM interferometers (see Fig.~\ref{fig:I3})
than in the case of local HOM interferometers (see Fig.~\ref{fig:I2}).

\section{Conclusions}
In this paper we studied two different sets of fundamental invariants 
of two-qubit states.
We demonstrated how to perform direct measurements of Makhlin's and 
Jing's et al. invariants by applying HOM interference on multiple copies
of the investigated two-qubit state. The developed techniques
for designing such interferometers can be useful for designing
new experiments for testing the quantum theory. We observed that
$W$-state projections needed in direct measurements of some of
the high order ($n=10,11,15,16,17,18$) invariants $I_n$ solve a classical problem 
of deciding handness  much faster than any classical (local) strategy.
Our analysis of Jing's et al. invariants also revealed that 
the nonlocal measurements or $W$-state projections 
are unnecessary for checking the equivalence of any two given
two-qubit states.
We demonstrated that by using nonlocal interferometers we learn 
the sign of $I_1$, which is not possible with using only local interferometers.
Learning this sign is an extra information gain appearing from
different (nonlocal) connections in the same quantum circuit.
This makes  the nonlocal measurements more efficient for
quantifying nonclassical correlations than the local ones in terms 
of the resources needed for such measurements.

Alice and Bob can learn the value of the sign of  $I_1$ only by
collaborating with each other, either by performing local or nonlocal measurements.
For this reason this extra information 
gain in the case of the joint measurements performed by Alice and Bob
could be useful in quantum information processing or communication
tasks similar to quantum secret sharing~\cite{Hillery99PRA},
but in a way that is invariant to local unitary operations.
Naturally, for testing fundamental physics of nonlocality one 
should perform only local measurements to avoid cyclic reference to
nonlocality. However, the nonlocal interferometers in some 
scenarios can be more useful for quantitative measurements.
We demonstrated the usefulness of nonlocal projections explicitly 
on the two examples of HOM interferometers designed to quantify 
CHSH nonlocality, linear entropy and fully-entangled fraction
with only local or both local and nonlocal
HOM interferometers. We compare these setups in context of
measuring nonlocality in Tab.~\ref{tab:summary}. 
Note that for all the method based on eigenvalues 
of $\hat R$ the product of the number of copies and the number of measurement
is constant and equals $12$. Thus, the effciency of these methods 
under perfect conditions would be the same and does not depend
on the number of copies. The method based directly on  
finding singular values of the correlation matrix $\hat \beta$ 
seems to be the most experimentally efficient but it is at the same time
the most mathematically complex. This means that the required calculations
can be computationally intensive and require some hardware to perform them. 
The singular values are typically found by first solving the eigenproblem for
$\hat \beta\hat \beta^T = \hat R$~\cite{riley2006mathematical}.
Thus, the presented  experimental methods based on eigenvalues of $\hat R$
can be interpreted as quantum-hardware implementations of
calculating functions of spectrum of $\hat \beta \hat \beta^T$
and they shift a part of the computational effort from postprocessing 
to the experiment. This gives a physical meaning to the abstract 
algebraic operations required for measuring
such fundamental quantities as nonclocality $M$ or 
fully-entangled fraction $f$, and other quantities defined 
via optimal  measurements. This implies 
the existence of a trade off between the experimental complexity
and computational complexity of the relevant measurements
and their postprocessing. 

\begin{table}
\caption{\label{tab:summary} 
The comparison of various methods for measuring 
Bell-CHSH nonlocality given by Eq.~(\ref{eq:M}).
Note that  one can use various combinations of numbers of
copies and measurements as some of the measurements can be performed in parallel
(see Figs.~\ref{fig:I1}-\ref{fig:I3}). 
 }
\begin{ruledtabular}
\begin{tabular}{cccc}
method &	copies &	measurements &	procedure \\
\hline
direct~\cite{CHSH69PRL} & 1 & $\infty$ & all CHSH inequalities\\
$\hat \beta$ matrix & 1 & 9 & local \\
$\hat R$ matrix & 2 & 6 & local \\
$I_{1},I_{2},I_{3}$ & 4 & 3 & nonlocal \\
$I_{1},I_{2},I_{3}$ & 6 & 2 & nonlocal \\
$I_{1},I_{2},I_{3}$ & 12 & 1 & nonlocal \\
$J_{1},J_{2},J_{3}$ & 6 & 2 & local \\
$J_{1},J_{2},J_{3}$ & 12 & 1 & local 
\end{tabular}
\end{ruledtabular}
\end{table}

\begin{acknowledgments}
We thank Adam Miranowicz for stimulating discussions. 
K.B. acknowledges the support by 
the Polish National Science Centre under grant No.
DEC-2013/11/D/ST2/02638 and support
by the Czech Science Foundation under the project No. 17-10003S.

\end{acknowledgments}

\appendix

\section{Detection events for the handness invariants}

\begin{figure*}
\includegraphics[width=0.75\linewidth]{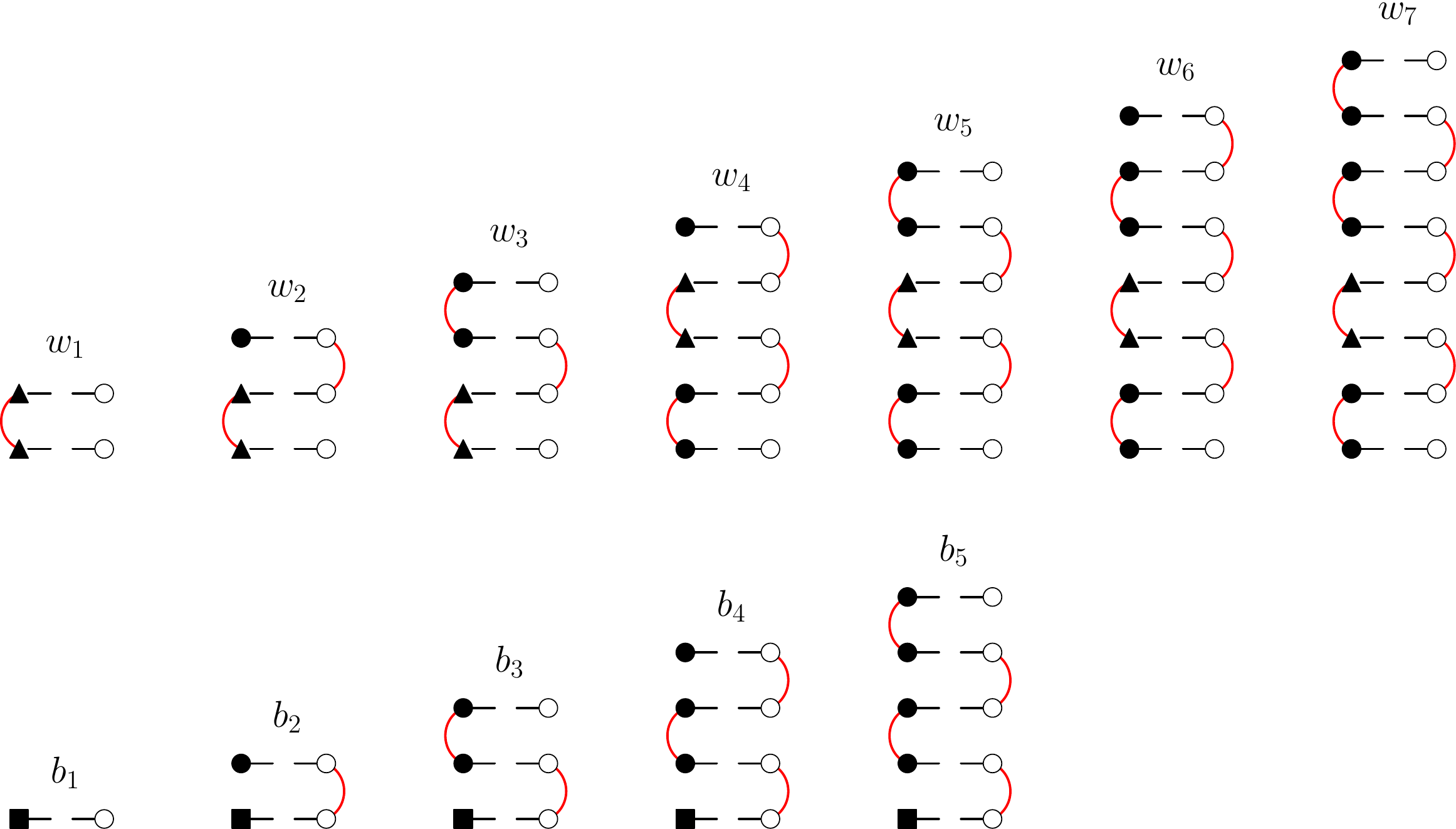}
\includegraphics[width=0.75\linewidth]{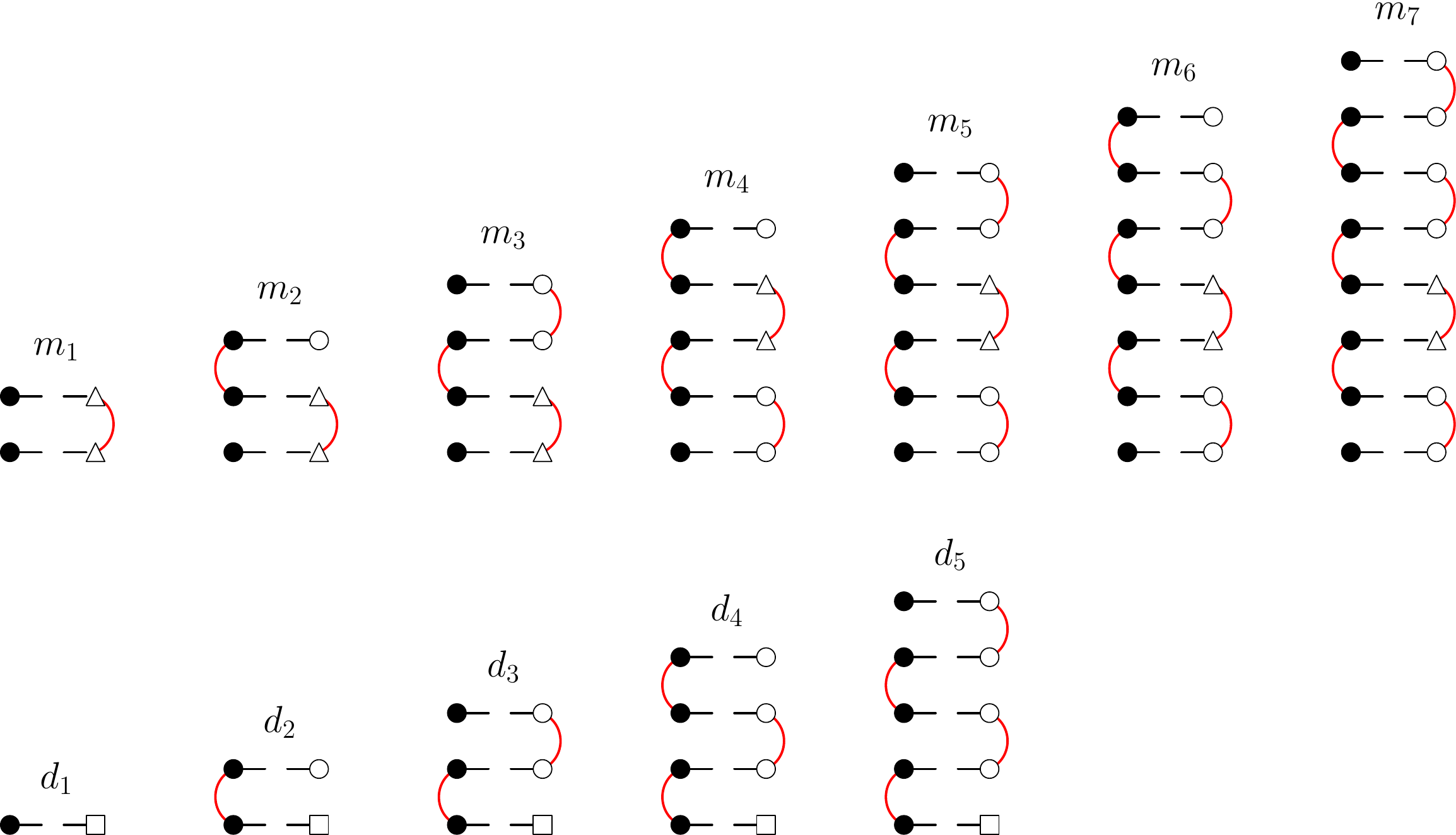}
\caption{\label{fig:Is}
The same as in Figs.~\ref{fig:chained} and~\ref{fig:looped}, but for
modified singlet projections (triangular markers) and $\hat\sigma_3$ measurements (square markers) involving $W$-sate projections
[see Eq.~(\ref{eq:Wklm})] needed for determining the handness invariants $I_n$ for $n=10,11,15,16,17,18$.}
\end{figure*}

Note that all the $W$-states used in this paper are invariant
under cyclic permutations. This fact can be used to group 
measurement outcomes. Our analysis of the expressions 
for Makhlin's invariants $I_n$ for  $n=10,11,15,16,17,18$
resulted in the complete list of the detection events
depicted in Fig.~\ref{fig:Is}.

\label{sec:appendix}


\bibliographystyle{apsrev4-1}


%

\end{document}